\documentclass[12pt,a4paper]{article}
\usepackage{graphicx}
\usepackage{cite}
\usepackage{amsmath}
\usepackage{myart}
\usepackage{amsfonts}
\usepackage{amssymb}
\newcommand\wtktN%
{{\kern-3.8pt}\widetilde{{\kern3.3pt}N{\kern3.3pt}}{\kern-3.4pt}\rangle}
\newcommand\wtbrN%
{\langle{\kern-3.4pt}\widetilde{{\kern3.3pt}N{\kern3.3pt}}{\kern-3.4pt}}

\begin{document}
\hfill SISSA 28/2007/EP

\bigskip

\begin{center}
{\Large{\bf On the $\boldsymbol{N{=}1}$ super Liouville four-point functions}}
\end{center}

\begin{center}
\vskip 1.0cm {\large
V.~A.~Belavin${}^{a,b}$}\\

\vspace{.4cm}
{\it
$^a$ International School for Advanced Studies (SISSA) \\
Via Beirut 2-4, 34014 Trieste, Italy \\
INFN sezione di Trieste \\
e-mail: belavinv@sissa.it

\vspace{0.3cm}
$^b$ Institute for Theoretical and Experimental Physics (ITEP)\\
B.~Cheremushkinskaya 25, 117259 Moscow, Russia
}
\end{center}

\vspace{1.0cm}

\textbf{Abstract:}
We construct the four-point correlation functions containing the top
component of the supermultiplet in the Neveu--Schwarz sector of the $N{=}1$
SUSY Liouville field theory. The construction is based on the recursive
representation for the NS conformal blocks. We test our results in the case
where one of the fields is degenerate with a singular vector on the level
$3/2$. In this case, the correlation function satisfies a third-order
ordinary differential equation, which we derive. We numerically verify the
crossing symmetry relations for the constructed correlation functions in the
nondegenerate case.

\section{Introduction}

The $N{=}1$ supersymmetric extension of the Liouville field theory
(SLFT)~\cite{Polyakov} plays an important role in a world-sheet description
of the noncritical fermionic string theory. It has at least two remarkable
properties. First, it is one of the simplest supersymmetric conformal field
theories (CFT); hence, investigating it might be useful for possible more
complicated generalizations. Second, it, in addition to the bosonic Liouville
theory, is another example of a ``noncompact" CFT, having a continuous
spectrum; this is a rather new object in the area of integrable models, worth
investigating in its own right. One of the main problems in any quantum field
theory is calculating correlation functions. Recent results obtained
in~\cite{SUSYboots} open the way for constructing the basic four-point
correlation functions in the NS sector of the SLFT. But there are still a few
subtleties in realizing this program. One is to classify the main correlation
functions clearly in terms of the basic superconformal block functions.
Another concerns the problem of finding the so-called elliptic uniformizing
representation for the four-point superconformal block functions. In this
paper, we partially tackle these problems.

We briefly recall the main features of the SLFT (see,
e.g.,~\cite{ZP, MSS, DHK, AG, Rubik} for more details). Local properties of
the SLFT are encoded in the Lagrangian density
\begin{equation}
\mathcal{L}_{\text{SLFT}}=\frac1{8\pi}(\partial_a\phi)^2+
\frac1{2\pi}(\psi\bar\partial\psi+\bar\psi\partial\bar\psi)+
2i\mu b^2\bar\psi\psi e^{b\phi}+2\pi b^2\mu^2e^{2b\phi}.
\label{SL}
\end{equation}
The symmetry is generated by the holomorphic and antiholomorphic components
of the supercurrent $S$ and the stress tensor $T$. In terms of the Laurent
components of $S$ and $T$, the algebra takes the conventional form of the
Neveu--Schwarz--Ramond~\cite{NS,R} algebra $SVir$:
\begin{equation}
\begin{aligned}
\lbrack L_n;L_m]&=(n-m)L_{n+m}+\frac{\hat c}8(n^3-n)\delta_{n;-m},
\\
\{G_r;G_s\}&=2L_{r+s}+\frac{\hat c}2\left(r^2-\frac14\right)\delta_{r;-s},
\\
[L_n;G_r]&=\left(\frac12n-r\right)G_{n+r},
\end{aligned}
\label{The algebra}
\end{equation}
where
\begin{equation*}
\begin{alignedat}{2}
&r,s\:\in\mathbb{Z}+\frac12&\quad&\text{for the NS sector},
\\
&r,s\:\in\mathbb{Z}&\quad&\text{for the R sector}.
\end{alignedat}
\end{equation*}
The central charge of the superconformal algebra is related to the parameter
$b$ in~\eqref{SL} via the ``background charge" $Q$,
\begin{equation}
\hat c=1+2Q^2,\quad\text{where }Q=b+\frac1b.
\end{equation}

Local fields in the SLFT belong to highest-weight representations of
$SVir\otimes\overline{SVir}$ algebra. Each representation, a so-called Verma
module, consists of a primary field $V_a$ with the conformal dimension
$\Delta=a(Q-a)/2$ (we sometimes use another convenient parameterization of
the conformal dimension $\Delta=Q^2/8-\lambda^2/2$) and all its
superconformal descendants. In this paper, we consider only the spinless
primary fields with $\bar\Delta=\Delta$. The general form of the descendant
operator is
\begin{equation}
\mathcal L_{\vec k}V_a=L_{-k_1}\cdots L_{-k_n}G_{-r_1}\cdots G_{-r_m}V_a,
\label{descendent}
\end{equation}
where $\vec k$ denotes $\{k_i,r_j\}$, which is an ordered set of positive
integers and half-integers correspondingly. The relation
$\sum_ik_i+\sum_jr_j=N$ fixes the particular level in the Verma module. It
is useful to introduce a special notation for the other components of the
primary supermultiplet:
\begin{equation}
\begin{aligned}
\Lambda_a&=G_{-1/2}V_a,
\\
\bar\Lambda_a&=\bar G_{-1/2}V_a,
\\
W_a&=G_{-1/2}\bar G_{-1/2}V_a.
\end{aligned}
\label{W}
\end{equation}

The space of fields in the SLFT forms an operator algebra closed under the
operator product expansion (OPE), which is continuous and involves
integration over the internal conformal dimension. The basic OPE is that of
two primary fields (for the sake of brevity below, we set
$\Delta=\Delta_{Q/2+iP}$ and $\Delta_i=\Delta_{a_i}$):
\begin{equation}
V_{a_1}(x)V_{a_2}(0)=\int\frac{dP}{4\pi}(x\bar x)^{\Delta-\Delta_1-\Delta_2}
\left(\mathbb{C}_{a_1,a_2}^{Q/2+iP}[V_{Q/2+iP}]_{\text{ee}}+
\mathbb{\widetilde C}_{a_1,a_2}^{Q/2+iP}[V_{Q/2+iP}]_{\text{oo}}\right).
\label{VV}
\end{equation}
The integration contour is basically along the real axis but should sometimes
be deformed (see~\cite{LFT}) under analytic continuation in the parameters
$a_1$ and $a_2$. In~\eqref{VV}, the ``chains" $[V_{Q/2+iP}]_{\text{ee}}$ and
$[V_{Q/2+iP}]_{\text{oo}}$ denote the respective contributions of integer and
half-integer descendents (the second subscript denotes the antiholomorphic
part; see Sec.~4). These contributions are unambiguously defined by the
superconformal invariance, but unlike the standard conformal symmetry, the
integer and half-integer descendents here enter independently. Hence, we
obtain two different structure constants $\mathbb{C}_{a_1a_2}^p$ and
$\mathbb{\widetilde C}_{a_1a_2}^p$. All other OPEs of two arbitrary local
fields can be derived from~\eqref{VV}. In particular,
\begin{align}
&W_{a_1}(x)V_{a_2}(0)=
\nonumber
\\
&\qquad\int\frac{dP}{4\pi}(x\bar x)^{\Delta-\Delta_1-\Delta_2-1/2}
\left(\mathbb{\widetilde C}_{a_1,a_2}^{Q/2+iP}
\widetilde{[V_{Q/2+iP}]}_{\text{ee}}+
\mathbb{C}_{a_1,a_2}^{Q/2+iP}\widetilde{[V_{Q/2+iP}]}_{\text{oo}}\right).
\label{WV}
\end{align}
We note that the structure constants are the same in both OPE~~\eqref{VV}
and~\eqref{WV}.

In this paper, we deal with the NS sector, which is closed under the OPE. In
the NS sector, the field $V_{m,n}$ with $m$ and $n$ being either both even or
both odd positive integers corresponds to the ``degenerate" primary field
with the conformal dimension $\Delta=\Delta(\lambda_{m,n})$,
\begin{equation}
\lambda_{m,n}=\frac{mb^{-1}+nb}2.
\end{equation}
The ``degenerate" primary field $V_{m,n}$ has a singular vector at the level
$N=mn/2$~\cite{Kac}. We introduce a ``singular-vector creation operator"
$D_{m,n}$~\cite{SUSYHEM} such that the singular vector appears when $D_{m,n}$
is applied to $V_{m,n}$. The normalization is fixed by taking the coefficient
of the leading term to be unity, $D_{m,n}=G_{-1/2}^{m,n}+\dots$. The first
nontrivial null vector in the NS sector occurs on the level $N=3/2$:
\begin{equation}
D_{13}V_{-b}=(G_{-1/2}^3+b^2G_{-3/2})V_{-b}=0.
\label{singvec}
\end{equation}

This paper is organized as follows. In Secs.~2 and~3, we consider the
four-point correlation functions in terms of the NS superconformal blocks and
analyze the special case where one field is the degenerate field $W_{-b}$. We
derive the corresponding differential equation, which becomes an important
tool in investigating the four-point correlation function in the subsequent
sections. In Sec.~4, we recapitulate the so-called recursive
``c-representation" and use it to construct the four-point correlation
function $\langle WVVV\rangle$. In Sec.~5, we suggest an elliptic
representation for the correlation function under consideration, making an
additional assumption about the asymptotic behavior of the superconformal
blocks considered as functions of the internal conformal dimension, and
numerically verify the crossing symmetry for the constructed four-point
correlation function. The last section contains a brief summary and some
discussion.

\section{Four-point correlation function and conformal blocks}

For our purposes, it makes sense to use the superfield formalism. Then the
four components of the supermultiplet can be joint in the primary superfield
$\widehat V_i=V_i+\theta_i\Lambda_i+\bar\theta_i\bar\Lambda_i+
\bar\theta_i\theta_iW$. In the supersymmetric case, in addition to the
standard anharmonic ratio $z$ for the four-point correlation functions, there
are two more independent superprojective invariants that should be taken into
account~\cite{ZP}. We let $\tau_1$ and $\tau_2$ denote them. The solution of
the superprojective constraints or, equivalently, the transformation law for
the primary superfields under superprojective transformations leads to the
general form of the the four-point correlation function
\begin{align}
&\langle\widehat V_{a_1}(z_1)\widehat V_{a_2}(z_2)\widehat V_{a_3}(z_3)
\widehat V_{a_4}(z_4)\rangle=
\nonumber
\\
&\qquad|z_{41}|^{-4\Delta_1}|z_{24}|^{2\Delta_{1+3-2-4}}
|z_{34}|^{2\Delta_{1+2-3-4}}|z_{23}|^{2\Delta_{4-1-2-3}}\widehat G
\left(\left.\begin{array}[c]{cc}%
a_1&a_2\\a_3&a_4\end{array}
\right|z,\bar z,\tau_1,\bar\tau_1,\tau_2,\bar\tau_2\right),
\label{anzac}
\end{align}
where $z_i=\{x_i,\theta_i\}$ is the holomorphic superspace coordinate and
$z_{ij}=x_i-x_j+\theta_i\theta_j$. Here, the reduced four-point function
$\widehat G$ depends only on superprojective invariants. Our choice of the
superprojective invariants is
\begin{equation}
\begin{aligned}
z&=\frac{z_{12}z_{34}}{z_{23}z_{41}},
\\
\tau_1&=\theta_{324},
\\
\tau_2&=z^{-1/2}(1-z)^{-1/2}\theta_{321},
\label{params1}
\end{aligned}
\end{equation}
where the ``odd" three-point superprojective invariant is
\begin{equation}
\theta_{ijk}=\frac{z_{jk}\theta_i+z_{ki}\theta_j+z_{ij}\theta_k-
\theta_i\theta_j\theta_k}{(z_{ij}z_{ik}z_{jk})^{1/2}}.
\end{equation}
For considering the correlation functions of the spinless fields, it suffices
to introduce four auxiliary functions $G_0$, $G_1$, $G_2$, and $G_3$:
\begin{align}
&\widehat G\left(\left.\begin{array}[c]{cc}%
a_1&a_2\\a_3&a_4\end{array}
\right|z,\tau_1,\bar\tau_1,\tau_2,\bar\tau_2\right)=
\nonumber
\\
&\qquad\qquad G_0(z,\bar z)+G_1(z,\bar z)\bar\tau_1\tau_1+
G_2(z,\bar z)\bar\tau_2\tau_2+G_3(z,\bar z)\bar\tau_1\tau_1\bar\tau_2\tau_2.
\label{G}
\end{align}
For the given choice of the superprojective invariants, the functions $G_i$
are uniquely related to the basic four-point correlation functions of the
components of primary supermultiplets. Our main focus in this paper is the
four-point correlation function involving one ``top" component $W$ of a
primary supermultiplet, while the others are ``bottom." For our choice of the
superprojective invariants,
\begin{equation}
\langle W_{a_1}(x)V_{a_2}(0)V_{a_3}(1)V_{a_4}(\infty)\rangle=
G_2\left(\left.\begin{array}[c]{cc}%
a_1&a_2\\a_3&a_4\end{array}\right|x,\bar x\right).
\label{G2}
\end{equation}
It can be evaluated via the integral representation based on OPEs~\eqref{VV}
and~\eqref{WV},
\begin{align}
&\langle W_{a_1}(x)V_{a_2}(0)V_{a_3}(1)V_{a_4}(\infty)\rangle=
\nonumber
\\
&\qquad\qquad\int\frac{dP}{4\pi}\mathbb{\widetilde C}_{a_1a_2}^{Q/2-iP}
\mathbb{C}_{a_3a_4}^{Q/2+iP}\mathcal{F}_{\text{e}}
\left(\left.\begin{array}[c]{cc}%
\hat a_1&a_3\\a_2&a_4\end{array}\right|\Delta,x\right)
\mathcal{F}_{\text{e}}\left(\left.\begin{array}[c]{cc}%
\hat a_1&a_3\\a_2&a_4\end{array}\right|\Delta,\bar x\right)
\nonumber
\\
&\quad\qquad+\int\frac{dP}{4\pi}\mathbb{C}_{a_1a_2}^{Q/2-iP}
\mathbb{\widetilde C}_{a_3a_4}^{Q/2+iP}\mathcal{F}_{\text{o}}
\left(\left.\begin{array}[c]{cc}%
\hat a_1&a_3\\a_2&a_4\end{array}\right|\Delta,x\right)
\mathcal{F}_{\text{o}}\left(\left.\begin{array}[c]{cc}%
\hat a_1&a_3\\a_2&a_4\end{array}\right|\Delta,\bar x\right),
\label{WVVV}
\end{align}
where $\mathcal{F}_{\text{e,o}}$ are the corresponding four-point
superconformal blocks with the intermediate dimension $\Delta=Q^2/4+P^2$
(the hat over $a_1$ highlights that the field is the ``top" component of the
supermultiplet). These functions were initially introduced in~\cite{BPZ}. In
Secs.~4 and~5, we present some details and explicit constructions concerning
this object. The basic NS structure constants in~\eqref{WVVV} were evaluated
in~\cite{Rubik, Marian} and were recently rederived without referring to the
Ramond sector~\cite{SUSYboots}:
\begin{equation}
\begin{aligned}
&\mathbb{C}^{Q-a_3}_{a_1a_2}=
\\
&\quad\left(\pi\mu\gamma\left(\frac{Qb}2\right)b^{1-b^2}\right)^{(Q-a)/b}
\frac{\Upsilon_{\text{NS}}'(0)\Upsilon_{\text{NS}}(2a_1)
\Upsilon_{\text{NS}}(2a_2)\Upsilon_{\text{NS}}(2a_3)}
{\Upsilon_{\text{NS}}(a_{1+2+3}-Q)\Upsilon_{\text{NS}}(a_{1+2-3})
\Upsilon_{\text{NS}}(a_{2+3-1})\Upsilon_{\text{NS}}(a_{3+1-2})},
\\
&\mathbb{\widetilde C}^{Q-a_3}_{a_1a_2}=
\\
&\quad-\left(\pi\mu\gamma\left(\frac{Qb}2\right)b^{1-b^2}\right)^{(Q-a)/b}
\frac{2i\Upsilon_{\text{NS}}'(0)\Upsilon_{\text{NS}}(2a_1)
\Upsilon_{\text{NS}}(2a_2)\Upsilon_{\text{NS}}(2a_3)}
{\Upsilon_{\text{R}}(a_{1+2+3}-Q)\Upsilon_{\text{R}}(a_{1+2-3})
\Upsilon_{\text{R}}(a_{2+3-1})\Upsilon_{\text{R}}(a_{3+1-2})},
\end{aligned}
\label{C3}
\end{equation}
where $a$ here denotes $a_1+a_2+a_3$ and we use the convenient notation 
in~\cite{Fukuda} for the special functions
\begin{equation}
\begin{aligned}
\Upsilon_{\text{NS}}(x)&=\Upsilon_b\left(\frac x2\right)
\Upsilon_b\left(\frac{x+Q}2\right),
\\
\Upsilon_{\text{R}}(x)&=\Upsilon_b\left(\frac{x+b}2\right)
\Upsilon_b\left(\frac{x+b^{-1}}2\right)
\end{aligned}
\label{YNSR}
\end{equation}
expressed in terms of the``upsilon" function $\Upsilon_b$ that is standard in
the Liouville field theory (see~\cite{DO, LFT}).

\section{Differential equations corresponding to the null vector~(1,3)}

We now turn to a four-point function with one singular field $W_{-b}$. It is
natural to renumber the operators, setting $W_{a_1}$ to be $W_{-b}$ and
respectively setting $V_{a_2}$, $V_{a_3}$, and $V_{a_4}$ to be $V_1$, $V_2$,
and $V_3$. As described in Appendix A, the decoupling of the singular vector
(1,3) leads to the partial differential equation for the four-point
correlation function
\begin{equation}
\left\{-b^{-2}\partial_4D_4+\sum_{i=1}^3\left\{
\frac{2\Delta_i}{z_{4i}^2}\theta_{4i}+\frac1{z_{4i}}\biggl[
2\theta_{4i}\partial_i-D_i\biggr]\right\}\right\}\langle{\widehat V}_{-b}(z_4)
{\widehat V}_1(z_1){\widehat V}_2(z_2){\widehat V}_3(z_3)\rangle=0,
\label{difur-1}
\end{equation}
which can be reduced to two systems of ordinary differential equations for
the holomorphic functions $g_0$, $g_1$, $g_2$, and $g_3$, which are just the
conformal blocks contributing to the functions $G_i$ introduced in~\eqref{G},
with the changes of the arguments described above. For example,
\begin{equation}
g_2(x)\Rightarrow G_2\left(\left.\begin{array}[c]{cc}%
-b&a_2\\a_1&a_3\end{array}\right|x,\bar x\right).
\end{equation}
The first system of equations is for $g_0$ and $g_3$,
\begin{equation}
\begin{aligned}
-b^{-2}xg_0''+\frac{3x-2}{x-1}g_0'+b^{-2}g_3'+
\bigg[\frac{\gamma_{13}}x+\frac{\gamma_{23}}{x-1}\bigg]g_0+
\frac{1-2x}{x(x-1)}g_3=0,&
\\
b^{-2}g_0''+\frac{1-3x}{x(x-1)}g_0'+\bigg[\frac{2\Delta_1}{x^2}+
\frac{2\Delta_2}{(x-1)^2}+\frac{2\gamma_{12}}{x(x-1)}\bigg]g_0+
\frac1{x(x-1)}g_3=0.&
\end{aligned}
\label{syst1}
\end{equation}
The second is for $g_1$ and $g_2$,
\begin{equation}
\begin{aligned}
-b^{-2}g_1''-b^{-2}xg_2''+\frac{3x-1}{x(x-1)}g_1'+
\bigg[b^{-2}+\frac{3x-2}{x-1}\bigg]g_2'-
\bigg[\frac{2\Delta_1}{x^2}+\frac{2\Delta_2}{(x-1)^2}&
\\
+\frac{3x-2}{x(x-1)^2}+\frac{2\gamma_{12}}{x(x-1)}\bigg]g_1+
\bigg[\frac{\gamma_{13}}x+\frac{\gamma_{23}}{x-1}-
\frac{6x^2-7x+3}{2x(x-1)^2}\bigg]g_2&=0,
\\
-b^{-2}g_2'+\frac{2x-1}{x(x-1)}g_2+\frac1{x(x-1)}g_1&=0,
\end{aligned}
\label{syst2}
\end{equation}
which leads to a third-order linear differential equation for $g_2$,
\begin{align}
&g_2'''+\frac{2(1-b^2)(1-2x)}{x(1-x)}g_2''
\nonumber
\\
&\qquad+\bigg(\frac{b^4-b^2+2 b^2(\Delta_1(1-x)+\Delta_2 x)}{x^2(1-x)^2}-
\frac{5b^4+b^2(2\Delta_3-7)+2}{x(1-x)}\bigg) g_2'
\nonumber
\\
&\qquad+b^4\bigg(\frac{3(\Delta_1-\Delta_2)+(b^2+\Delta_3-1)(1-2x)}
{x^2(1-x)^2}+\frac{2\Delta_2x-2\Delta_1(1-x)}{x^3(1-x)^3}\bigg)g_2=0.
\label{g2}
\end{align}
Three independent solutions for $g_2$ with a diagonal monodromy near $x=0$
are just the $s$-channel conformal blocks corresponding to $G_2$. Using the
definitions of the conformal blocks (see~\eqref{ConfBlockDef2} below) to
compare the exponents for $x\to0$, we find
\begin{equation}
\begin{aligned}
&g_2^{(\pm)}=x^{(1+b^2\pm2b\lambda_1)/2}\sum_{n=0}^{\infty}
A_n^{(\pm)}x^n=\mathcal{F}_{\text{o}}\left(\left.\begin{array}[c]{cc}%
\hat\Delta_{13}&\Delta_3\\ \Delta_1&\Delta_4\end{array}
\right|\Delta(\lambda_1\mp b)\bigg|x\right),
\\
&g_2^{(0)}=x^{b^2}\sum_{n=0}^{\infty}A_n^{(0)}x^n=
\mathcal{F}_{\text{e}}\left(\left.\begin{array}[c]{cc}%
\hat\Delta_{13}&\Delta_3\\ \Delta_1&\Delta_4\end{array}
\right|\,\Delta(\lambda_1)\,\bigg|x\right),
\end{aligned}
\label{g20+-}
\end{equation}
corresponding to the overall normalization
\begin{equation}
\begin{aligned}
&A_0^{(0)}=1,
\\
&A_0^{(\pm)}=-\frac{\Delta(\lambda_1\mp b)+\Delta_{13}-
\Delta(\lambda_1)}{2\Delta(\lambda_1\mp b)}.
\end{aligned}
\label{norm}
\end{equation}
The first terms in the series expansion can be easily found by substituting
these expansions in differential equation~\eqref{g2} and solving the
recursive relations for the coefficients order by order. For example, for the
solution $g_2^{(0)}$ and $g_2^{(+)}$, we have
\begin{align}
A_0^{(0)}={}&1,
\\
A_1^{(0)}={}&\bigl(b^2(-1-2 b^2-b^4+4b^2\lambda_1^2+4b^2\lambda_2^2-
4b^2\lambda_3^2)\bigl)
\nonumber
\\&{}\times\bigl(2(1+b^2-2b\lambda_1)(1+b^2+2b\lambda_1)\bigr)^{-1},
\\
A_2^{(0)}={}&(b^2(-22-47b^2-24b^4+6b^6+6b^8+b^{10}+96b^2\lambda_1^2+
24b^4\lambda_1^2-16b^6\lambda_1^2-8b^8\lambda_1^2
\nonumber
\\
&{}-32b^4\lambda_1^4+16b^6\lambda_1^4+80b^2\lambda_2^2-8b^4\lambda_2^2-
32b^6\lambda_2^2-8b^8\lambda_2^2-64b^4\lambda_1^2\lambda_2^2+
32b^6\lambda_1^2\lambda_2^2\qquad
\nonumber
\\
&{}+32b^4\lambda_2^4+16b^6\lambda_2^4-64b^2\lambda_3^2+40b^4\lambda_3^2+
48b^6\lambda_3^2+8b^8\lambda_3^2-32b^6\lambda_1^2\lambda_3^2
\nonumber
\\
&{}-64b^4\lambda_2^2\lambda_3^2-32b^6\lambda_2^2\lambda_3^2+
32b^4\lambda_3^2+16b^6\lambda_3^4)\bigr)
\bigl(8(1+b^2-2b\lambda_1)(3+b^2-2b\lambda_1)
\nonumber
\\
&{}\times(1+b^2+2b\lambda_1)(3+b^2+2b\lambda_1)\bigr)^{-1},
\\
A_0^{(+)}={}&2b^2(-1+b^2-2b\lambda_1)^{-1},
\\
A_1^{(+)}={}&b^2(3-10b^2+3b^4+8b\lambda_1-8b^3\lambda_1+4b^2\lambda_1^2
+4b^2\lambda_2^2-4b^2\lambda_3^2)
\nonumber
\\
&{}\times\bigl(2(1-b^2+2b\lambda_1)(-3+b^2-2b\lambda_1)\bigr)^{-1}.
\end{align}
These expressions coincide with the corresponding
terms~\eqref{F0}--\eqref{F1} of the conformal block series expansions
calculated for the same parameter choices and thus confirm the recursive
relations for the conformal blocks discussed in the next sections.

To conclude this section, we analyze the solutions of differential
equation~\eqref{g2} and their monodromy properties. We show that these
properties lead to functional relations for the structure constants that are
consistent with~\eqref{C3}. The decoupling equation restricts
OPEs~\eqref{VV} and~\eqref{WV} to the ``discrete" form (cf.~\eqref{g20+-})
\begin{equation}
\begin{aligned}
&V_{-b}(x)V_a(0)=
\\
&\qquad(x\bar x)^{ab}C_-(a)[V_{a-b}]_{\text{ee}}+
(x\bar x)^{1/2+b^2}\widetilde C_0(a)[V_a]_{\text{oo}}+
C_+(a)(x\bar x)^{1-ba+b^2}[V_{a+b}]_{\text{ee}},
\\
&W_{-b}(x)V_a(0)=
\\
&\qquad(x\bar x)^{ab}C_-(a)\widetilde{[V_{a-b}]}_{\text{oo}}+
(x\bar x)^{b^2}\widetilde C_0(a)\widetilde{[V_a]}_{\text{ee}}+
C_+(a)(x\bar x)^{1-ba+b^2}\widetilde{[V_{a+b}]}_{\text{oo}},
\end{aligned}
\label{degOPE}
\end{equation}
where the special structure constants are
\begin{equation}
\begin{aligned}
&C_-(a)=\operatorname*{res}_{\epsilon=0}
\mathbb{C}_{a,-b+\epsilon}^{a-b-\epsilon}=1,
\\
&C_0(a)=\operatorname*{res}_{\epsilon=0}
\mathbb{\widetilde C}_{a,-b+\epsilon}^{a-\epsilon}=
\frac{2\pi i\mu}{\gamma(-b^2)\gamma(ba_1)\gamma(1+b^2-ba_1)},
\\
&C_+(a)=\operatorname*{res}_{\epsilon=0}
\mathbb{C}_{a,-b+\epsilon}^{a+b-\epsilon}=
\left(\gamma\left(\frac{Qb}2\right)\right)^2\,
\frac{(\pi\mu)^2b^4\gamma(a_1b-1/2-b^2/2)}{\gamma(1/2+b^2/2+a_1b)}.
\end{aligned}
\end{equation}
These expressions can be found (see, e.g.,~\cite{SUSYboots}) directly from
SLFT Lagrangian~\eqref{SL} by using perturbative calculations with the
screening operator, similar to that developed in~\cite{DF}. With general
expression~\eqref{WVVV} and OPE~\eqref{degOPE} taken into account,
the correlation function $G_2$ is combined as
\begin{align}
G_2\left(\left.\begin{array}[c]{cc}%
-b&a_2\\a_3&a_4\end{array}\right|x,\bar x\right)={}&
C_-(a_1)\widetilde C_{a_1-b,a_2,a3}
\mathcal{F}_{\text{o}}^{(-)}(x)\mathcal{F}_{\text{o}}^{(-)}(\bar x)
\nonumber
\\&{}+C_+(a_1)\widetilde C_{a_1+b,a_2,a3}
\mathcal{F}_{\text{o}}^{(+)}(x)\mathcal{F}_{\text{o}}^{(+)}(\bar x)
\nonumber
\\
&+\widetilde C_0(a_1)C_{a_1,a_2,a3}
\mathcal{F}_{\text{e}}^{(0)}(x)\mathcal{F}_{\text{e}}^{(0)}(\bar x),
\label{g2a}
\end{align}
On the other hand, substituting
\begin{equation}
g_2(x)=x^{a_1b}(1-x)^{a_2b}F(x)
\label{gF}
\end{equation}
converts~\eqref{g2} to the form
\begin{align}
&x^2(1-x)^2F'''-x(1-x)(K_1x-K_2(1-x))F''
\nonumber
\\
&\qquad{}+(L_1x^2+L_2(1-x)^2-L_3x(1-x))F'+(M_1x-M_2(1-x))F=0.
\label{FDeq}
\end{align}
It turns out that this differential equation (as well as the one recently
considered for the correlation function $g_0$~\cite{SUSYboots}) is of a
special type~\cite{DF} and can be solved in terms of two-fold contour
integrals of the form
\begin{equation}
{\displaystyle\int\limits_{C_{\alpha}}}
{\displaystyle\int\limits_{C_{\beta}}}
dt_1\,dt_2\,| t_1t_2|^A|(1-t_1)(1-t_2)|^B|(x-t_1)(x-t_2)|^C|t_1-t_2|^{2g}.
\label{Iab}
\end{equation}
For ansatz~\eqref{Iab} to be the solution of differential
equation~\eqref{FDeq}, the exponents $A$, $B$, $C$, and $g$ must satisfy the
system of constraints
\begin{equation}
\begin{aligned}
&\begin{alignedat}{2}
K_1&=-2g-3B-3C,&\qquad K_2&=-2g-3A-3C,
\\
L_1&=(B+C)(2B+2C+2g+1),&\qquad L_2&=(A+C)(2A+2C+2g+1),
\end{alignedat}
\\
&\begin{aligned}
L_3&=4AB+4(2A+2B+2C+1)C+4(A+B+3C)g+4g^2+2g,
\\
M_1&=-2C(A+B+C+g+1)(2B+2C+2g+1),
\\
M_2&=-2C(A+B+C+g+1)(2A+2C+2g+1),
\end{aligned}
\end{aligned}
\label{KLM}
\end{equation}
which coincides with the one that appeared in~\cite{SUSYboots} but with
different values of $K_i$, $L_i$, and $M_i$. This system is overdetermined
for an arbitrary choice of $K_i$, $L_i$, and $M_i$. But it can be solved in
our case and yields the following expressions for $A$, $B$, $C$, and $g$ in
terms of the basic parameters $a_1$, $a_2$, $a_3$, and $b$:
\begin{equation}
\begin{alignedat}{2}
A&=-1+\frac{ba_{2+3-1}}2,&\qquad B&=-1+\frac{ba_{1-2+3}}2,
\\
C&=-\frac{ba_{1+2+3}}2+b^2,&\qquad g&=\frac12-\frac{b^2}2.
\end{alignedat}
\label{ABC}
\end{equation}
Hence, the analogous consideration in~\cite{SUSYboots} is also applicable in
this case. In particular, the single-valued solution of Eq.~\eqref{g2} (and
of the same equation with respect to $\bar x$) can be constructed as
\begin{equation}
g_2(x,\bar x)=X_1\mathcal{I}_1(x)\mathcal{I}_1(\bar x)+X_2\mathcal{I}_2(x)
\mathcal{I}_2(\bar x)+X_3\mathcal{I}_3(x)\mathcal{I}_3(\bar x).
\label{g2b}
\end{equation}
Here, $\mathcal{I}_i(x)$ are the three independent solutions with a diagonal
monodromy around $x=0$
\begin{equation}
\begin{aligned}
\mathcal{I}_1(x)&=\mathcal{I}_1^{(0)}(1+\dots),
\\
\mathcal{I}_2(x)&=x^{1+A+C}\mathcal{I}_2^{(0)}(1+\dots),
\\
\mathcal{I}_3(x)&=x^{2+2A+2C+2g}\mathcal{I}_3^{(0)}(1+\dots),
\end{aligned}
\label{I123}
\end{equation}
where the dots denote a regular series in $x$ and
\begin{equation}
\begin{aligned}
\mathcal{I}_1^{(0)}&=
\frac{\Gamma(2g)\Gamma(1\,{+}\,B)\Gamma(1\,{+}\,B\,{+}\,g)
\Gamma(-1\,{-}\,2g\,{-}\,A\,{-}\,B\,{-}\,C)
\Gamma(-1\,{-}\,g\,{-}\,A\,{-}\,B\,{-}\,C)}
{\Gamma(g)\Gamma(-g-A-C)\Gamma(-A-C)},\quad
\\
\mathcal{I}_2^{(0)}&=
\frac{\Gamma(1+A)\Gamma(1+B)\Gamma(1+C)\Gamma(-1-2g-A-B-C)}
{\Gamma(2+A+C)\Gamma(-2g-A-C)},
\\
\mathcal{I}_3^{(0)}&=
\frac{\Gamma(2g)\Gamma(1+A)\Gamma(1+A+g)\Gamma(1+C)\Gamma(1+C+g)}
{\Gamma(g)\Gamma(2+A+C+g)\Gamma(2+A+C+2g)}.
\end{aligned}
\label{Iconst}
\end{equation}
The function $g_2(x,\bar x)$ coincides with correlation function~\eqref{g2a}
up to overall normalization. It was established in~\cite{FD2} that
\begin{equation}
\begin{aligned}
\frac{X_3}{X_1}&=
\frac{\sin\pi A\sin\pi C\sin\pi(A+C)\sin\pi(A+g)\sin\pi(C+g)}
{\sin\pi B\sin\pi(B{+}g)\sin\pi(A{+}B{+}C{+}g)
\sin\pi(A{+}C{+}2g)\sin\pi(A{+}B{+}C{+}2g)},
\\
\frac{X_2}{X_1}&=\frac{\sin\pi(A+C+g)\sin\pi A\sin\pi C}
{2\cos\pi g\sin\pi(B+g)\sin\pi(A+B+C+g)\sin\pi(A+C+2g)}.
\end{aligned}
\label{XX}
\end{equation}
Comparing expressions~\eqref{g2a} and~\eqref{g2b} (with the normalization of
the conformal blocks encoded in~\eqref{VV} and~\eqref{WV} and resulting
in~\eqref{norm} taken into account) results in the relations for the
structure functions
\begin{align}
&{\kern26mm}
\frac{C_-(a_1)\widetilde C_{a_1-b,a_2,a_3}}
{\widetilde C_0(a_1)C_{a_1,a_2,a_3}}=
\frac{(a_1-b)^2X_1{\mathcal{I}_1^{(0)}}^2}{b^2X_2{\mathcal{I}_2^{(0)}}^2},
\\
&\frac{\widetilde C_{a_1-b,a_2,a_3}}{C_{a_1,a_2,a_3}}=
2\pi i\mu\frac{\gamma(\frac{1+b^2}2)\gamma(\frac{a_{1+2+3}}{2}-b^2)
\gamma(\frac{1+ba_{1-2+3}-b^2}{2})\gamma(\frac{1+ba_{1+2-3}-b^2}{2})}
{\gamma(ba_1-b^2)\gamma(\frac{ba_{-1+2+3}}{2})\gamma(\frac{1+ba_1-b^2}2)}.
\label{CCa}
\end{align}
Similarly,
\begin{align}
&{\kern20mm}
\frac{C_+(a_1)\widetilde C_{a_1+b,a_2,a_3}}
{C_-(a_1)\widetilde C_{a_1-b,a_2,a_3}}=
\frac{(a_1-b^{-1})^2X_3{\mathcal{I}_3^{(0)}}^2}
{(a_1-b)X_1 {\mathcal{I}_1^{(0)}}^2},
\\
&\frac{\widetilde C_{a_1+b,a_2,a_3}}{\widetilde C_{a_1-b,a_2,a_3}}=
\frac{\gamma(ba_1)\gamma(ba_1-b^2)\gamma(\frac{1+2b a_1-b^2}2)
\gamma(\frac{1+2b a_1+b^2}2)}{\pi^2\mu^2b^4\gamma^2(\frac{1+b^2}2)}
\nonumber
\\
&\phantom{\frac{\widetilde C_{a_1+b,a_2,a_3}}
{\widetilde C_{a_1-b,a_2,a_3}}={}}
\times\frac{\gamma(\frac{b a_{-1+2+3}}2)}
{\gamma(\frac{ba_{1+2-3}}{2})\gamma(\frac{ba_{1-2+3}}{2})
\gamma(\frac{ba_{1+2+3}}{2}-b^2)}
\nonumber
\\
&\phantom{\frac{\widetilde C_{a_1+b,a_2,a_3}}
{\widetilde C_{a_1-b,a_2,a_3}}={}}
\times\frac{\gamma(\frac{1+ba_{-1+2+3}-b^2}2)}
{\gamma(\frac{1+ba_{1-2+3}-b^2}{2})\gamma(\frac{-1+b a_{1+2+3}-b^2}{2})
\gamma(\frac{1+b a_{1+2-3}-b^2}{2})}.
\label{CCb}
\end{align}
It can be verified directly that structure constants~\eqref{C3} satisfy
functional relations~\eqref{CCa} and~\eqref{CCb}.

\section{Analytic properties of the conformal blocks and the
$\hat c$-recursion}

Taking the superprojective invariance into account, we can see that there are
only four independent spinless four-point correlation functions
(see~\eqref{G}). The other twelve spinless correlation functions are related
to these four by superprojective transformations. We take the correlators
$\langle VVVV\rangle$, $\langle WVVV\rangle$, $\langle VVWV\rangle$, and
$\langle WVWV\rangle$ as our basis four-point functions. The correlation
function $\langle VVVV\rangle$ of the ``bottom" components of the
supermultiplet was considered recently (see~\cite{vbelavin, leshek} for the
details). In this paper, we are interested in the correlation function
$\langle WVVV\rangle$. The idea for evaluating the conformal blocks
contributing to~\eqref{WVVV} is similar to that in the previous case. The
necessery $s$-channel superconformal blocks are defined via the expansions
\begin{align}
&\mathcal{F}_{\text{e}}\left(\left.\begin{array}[c]{cc}%
\hat a_1&a_3\\a_2&a_4\end{array}\right|\Delta\,\bigg|\,x\right)=
x^{\Delta-\Delta_1-\Delta_2-1/2}\sum^{N\text{ integer}}_{N\ge0}x^N{}_{12}
\wtbrN|N\rangle_{34},
\label{ConfBlockDef1}
\\
&\mathcal{F}_{\text{o}}\left(\left.\begin{array}[c]{cc}%
\hat a_1&a_3\\a_2&a_4\end{array}\right|\Delta\,\bigg|\,x\right)=
x^{\Delta-\Delta_1-\Delta_2-1/2}\sum^{N\text{ half-integer}}_{N>0}x^N{}_{12}
\wtbrN|N\rangle_{34},
\label{ConfBlockDef2}
\end{align}
where the ``chain" vectors $|N\rangle$ and $|\wtktN$
are the respective $N$th-level descendent contributions of the intermediate
state with the conformal dimension $\Delta$ appearing in OPEs~\eqref{VV}
and~\eqref{WV} for $V(x)V(0)$ and $W(x)V(0)$,
\begin{equation}
\begin{aligned}
&[V_{\Delta}]_{\text{ee,oo}}=
\sum_{N\in\mathbb{Z},\mathbb{Z}/2}x^N|N\rangle,
\\
&\widetilde{[V_{\Delta}]}_{\text{ee,oo}}=
\sum_{N\in\mathbb{Z},\mathbb{Z}/2}x^N|\wtktN.
\end{aligned}
\end{equation}
Here, we suppress the dependence of the chain operators on the external
dimensions. We introduce the operators $\mathcal{C}(N,\Delta)$ creating the
chain vector $|N\rangle=\mathcal{C}(N,\Delta)V_{\Delta}$ and the similar
operator $\mathcal{\widetilde C}(N,\Delta)$ for the chain vector
$|\wtktN$. The vectors $|N\rangle$ and
$|\wtktN$ are completely determined by the
superconformal symmetry leading to the following relations for the vectors of
the chain that grows from the vacuum vector $V_{\Delta}$:
\begin{equation}
\begin{aligned}
G_k\mathcal{C}(N,\Delta)V_{\Delta}&=
\mathcal{\widetilde C}(N-k,\Delta)V_{\Delta}
\\
G_k\mathcal{\widetilde C}(N,\Delta)V_{\Delta}&=
(\Delta+2\Delta_1k-\Delta_2+N-k)\mathcal{C}(N-k,\Delta)V_{\Delta}
\end{aligned}
\label{chain}
\end{equation}
for $0<k\le N$.

We note that the conformal blocks contributing to the other two basis
correlation functions are also expressed in terms of the chain vectors
$|N\rangle$ and $|\wtktN$. Namely, the superconformal
blocks have the forms
\begin{align}
&\mathcal{F}_{\text{e}}\left(\left.\begin{array}[c]{cc}%
a_1&\hat a_3\\a_2&a_4\end{array}\right|\Delta\,\bigg|\,x\right)=
x^{\Delta-\Delta_1-\Delta_2}\sum^{N\text{ integer}}_{N\ge0}
x^N{}_{12}\langle N|\wtktN_{34},
\label{ConfBlockDef1a}
\\
&\mathcal{F}_{\text{o}}\left(\left.\begin{array}[c]{cc}%
a_1&\hat a_3\\a_2&a_4\end{array}\right|\Delta\,\bigg|\,x\right)=
x^{\Delta-\Delta_1-\Delta_2}\sum^{N\text{ half-integer}}_{N>0}
x^N{}_{12}\langle N|\wtktN_{34}
\label{ConfBlockDef3}
\end{align}
for the correlation function $\langle VVWV\rangle$ and
\begin{align}
&\mathcal{F}_{\text{e}}\left(\left.\begin{array}[c]{cc}%
\hat a_1&\hat a_3\\a_2&a_4\end{array}\right|\Delta\,\bigg|\,x\right)=
x^{\Delta-\Delta_1-\Delta_2-1/2}\sum^{N\text{ integer}}_{N\ge0}
x^N{}_{12}\wtbrN|\wtktN_{34},
\label{ConfBlockDef1b}
\\
&\mathcal{F}_{\text{o}}\left(\left.\begin{array}[c]{cc}%
\hat a_1&\hat a_3\\a_2&a_4\end{array}\right|\Delta\,\bigg|\,x\right)=
x^{\Delta-\Delta_1-\Delta_2-1/2}\sum^{N\text{ half-integer}}_{N>0}
x^N{}_{12}\wtbrN|\wtktN_{34}
\label{ConfBlockDef3a}
\end{align}
for the correlation function $\langle WVWV\rangle$.

Below, instead of speaking in terms of the chain vectors $|N\rangle$ and
$|\wtktN$, we use the chain operators~\cite{SUSYboots}
defined as
\begin{equation}
\begin{aligned}
\mathcal{C}(\Delta,x)&=\mathcal{C}_{\text{e}}(\Delta,x)+
\mathcal{C}_{\text{o}}(\Delta,x),
\\
\mathcal{\widetilde C}(\Delta,x)&=
\mathcal{\widetilde C}_{\text{e}}(\Delta,x)+
\mathcal{\widetilde C}_{\text{o}}(\Delta,x),
\end{aligned}
\label{chains}
\end{equation}
where
\begin{equation}
\mathcal{C}_{\text{e,o}}(\Delta,x)=
\sum_{N\in\mathcal{Z},\mathcal{Z}/2}x^N\mathcal{C}(N,\Delta),\qquad
\mathcal{\widetilde C}_{\text{e,o}}(\Delta,x)=
\sum_{N\in \mathcal{Z},\mathcal{Z}/2}x^N\mathcal{\widetilde C}(N,\Delta).
\label{Cn}
\end{equation}
The consideration in terms of chain operators~\eqref{chains} allows
presenting the results more transparently. The analytic properties of the
chain operators were discussed previously~\cite{SUSYboots, vbelavin}. Here,
we present only the main results to make the material more independent. It
follows from the properties of Eqs.~\eqref{chain} that starting from the
level $mn/2$, the chain operators have simple poles at $\Delta=\Delta_{m,n}$,
and the residues at these poles are proportional to the new chain
$\mathcal{C}(\Delta_{m,-n},x)D_{m,n}V_{m,n}$ or
$\mathcal{\widetilde C}(\Delta_{m,-n},x)D_{m,n}V_{m,n}$ growing from the
vector $D_{m,n}V_{\Delta}$, which themselves satisfy chain
equations~\eqref{chain} with $\Delta_{m,n}$ substituted for $\Delta$. We have
the residue formulas
\begin{equation}
\begin{aligned}
\operatorname*{res}_{\Delta=\Delta_{m,n}}\mathcal{C}_{\text{e}}(\Delta,x)&=
x^{mn/2}\begin{cases}
X_{m,n}^{\text{(e)}}\mathcal{C}_{\text{e}}(\Delta_{m,-n},x)D_{m,n},
&m,n\text{ even},\\
X_{m,n}^{\text{(e)}}\mathcal{C}_{\text{o}}(\Delta_{m,-n},x)D_{m,n},
&m,n\text{ odd},\end{cases}
\\
\operatorname*{res}_{\Delta=\Delta_{m,n}}\mathcal{C}_{\text{o}}(\Delta,x)&=
x^{mn/2}\begin{cases}
X_{m,n}^{\text{(o)}}\mathcal{C}_{\text{o}}(\Delta_{m,-n},x)D_{m,n},
&m,n\text{ even},\\
X_{m,n}^{\text{(o)}}\mathcal{C}_{\text{e}}(\Delta_{m,-n},x)D_{m,n},
&m,n\text{ odd},
\end{cases}
\end{aligned}
\label{resC}
\end{equation}
for the chain $\mathcal{C}(\Delta,x)$ and
\begin{equation}
\begin{aligned}
\operatorname*{res}_{\Delta=\Delta_{m,n}}
\mathcal{\widetilde C}_{\text{e}}(\Delta,x)&=x^{mn/2}\begin{cases}
X_{m,n}^{\text{(o)}}\mathcal{C}_{\text{e}}(\Delta_{m,-n},x)D_{m,n},
&m,n\text{ even},\\
X_{m,n}^{\text{(o)}}\mathcal{C}_{\text{o}}(\Delta_{m,-n},x)D_{m,n},
&m,n\text{ odd},\end{cases}
\\
\operatorname*{res}_{\Delta=\Delta_{m,n}}
\mathcal{\widetilde C}_{\text{o}}(\Delta,x)&=x^{mn/2}\begin{cases}
X_{m,n}^{\text{(e)}}\mathcal{C}_{\text{o}}(\Delta_{m,-n},x)D_{m,n},
&m,n\text{ even},\\
X_{m,n}^{\text{(e)}}\mathcal{C}_{\text{e}}(\Delta_{m,-n},x)D_{m,n},
&m,n\text{ odd},\end{cases}
\end{aligned}
\label{resCt}
\end{equation}
for the complementary chain $\mathcal{\widetilde C}(\Delta,x)$. Here,
\begin{equation}
X_{m,n}^{\text{(e,o)}}(\Delta_1,\Delta_2)=2^{-p_{\text{e,o}}(m,n)}
\frac{P_{m,n}^{\text{(e,o)}}(\lambda_1+\lambda_2)P_{m,n}^{\text{(e,o)}}
(\lambda_2-\lambda_1)}{r_{m,n}'},
\label{YPP}
\end{equation}
where the degree $p_{\text{e,o}}(m,n)=\deg P_{m,n}^{\text{(e,o)}}(x)$ of the
``fusion polynomials"~\cite{vbelavin, leshek} in the numerator,
\begin{equation}
\begin{aligned}
P_{m,n}^{\text{(e)}}(x)&=
\prod_{\substack{k\in\{1-m,2,m-1\}\\l\in\{1-n,2,n-1\}}}%
^{m+n-k-l\operatorname*{mod}4=0}(x-\lambda_{k,l}),
\\
P_{m,n}^{\text{(o)}}(x)&=
\prod_{\substack{k\in\{1-m,2,m-1\}\\l\in\{1-n,2,n-1\}}}%
^{m+n-k-l\operatorname*{mod}4=2}(x-\lambda_{k,l})
\end{aligned}
\label{Peo}
\end{equation}
(the notation $\{1-m,2,m-1\}$, for example, means ``from $1-m$ to $m-1$ with
step 2," i.e., $1-m,3-m,\dots,m-1$), coincides with the number of multipliers
in products~\eqref{Peo},
\begin{equation}
\begin{aligned}
&\,\,\,p_{\text{e,o}}(m,n)=mn/2\quad\text{for }m,n\text{ even},
\\
&\left.\begin{aligned}
&p_{\text{e}}(m,n)=mn/2-1/2\\
&p_{\text{o}}(m,n)=mn/2+1/2\end{aligned}\right\}
\quad\text{for }m,n\text{ odd}.
\end{aligned}
\label{pmn}
\end{equation}
The denominator in~\eqref{YPP} is related to the norm of the ``quasisingular"
vector $D_{m,n}V_{\Delta}$, which was explicitly evaluated in~\cite{SUSYHEM},
\begin{equation}
r_{m,n}'=2^{mn-1}\prod_{\substack{k=1-m,\,l=1-n\\
(k,l)\ne(0,0),(m,n)}}^{m,\,n,\,k+l\in2\mathbb{Z}}\lambda_{k,l}.
\label{rp}
\end{equation}
These simple analytic properties are inherited by superconformal
blocks~\eqref{ConfBlockDef1} and~\eqref{ConfBlockDef2}. The poles of
$\mathcal{C}_{\text{e,o}}(\Delta)V_{\Delta}$ become the poles of the blocks,
the residues being evaluated similarly (we suppress the external dimensions
in the arguments of the blocks for compactness)
\begin{equation}
\begin{aligned}
\operatorname*{res}_{\Delta=\Delta_{m,n}}
\mathcal{F}_{\text{e}}(\Delta,x)&=
B_{m,n}^{\text{(e)}}\left(\begin{array}[c]{cc}%
\hat\Delta_1&\Delta_3\\ \Delta_2&\Delta_4\end{array}\right)\begin{cases}
\mathcal{F}_{\text{e}}(\Delta_{m,-n},x)&\text{for }m,n\text{ even},\\
\mathcal{F}_{\text{e}}(\Delta_{m,-n},x)&\text{for }m,n\text{ odd},\end{cases}
\\
\operatorname*{res}_{\Delta=\Delta_{m,n}}
\mathcal{F}_{\text{o}}(\Delta,x)&=
B_{m,n}^{\text{(o)}}\left(\begin{array}[c]{cc}%
\hat\Delta_1&\Delta_3\\ \Delta_2&\Delta_4\end{array}\right)\begin{cases}
\mathcal{F}_{\text{o}}(\Delta_{m,-n},x)&\text{for }m,n\text{ even},\\
\mathcal{F}_{\text{e}}(\Delta_{m,-n},x)&\text{for }m,n\text{ odd},\end{cases}
\end{aligned}
\label{resB}
\end{equation}
where
\begin{equation}
B_{m,n}^{\text{(e,o)}}\left(\begin{array}[c]{cc}%
\Delta_1&\Delta_3\\ \Delta_2&\Delta_4\end{array}\right)=
r_{m,n}' X_{m,n}^{\text{(o,e)}}(\Delta_1,\Delta_2)
X_{m,n}^{\text{(e,o)}}(\Delta_3,\Delta_4).
\label{Beo}
\end{equation}
It follows from~\eqref{resB} that conformal blocks~\eqref{ConfBlockDef1}
and~\eqref{ConfBlockDef2} as functions of the central charge $\hat c$ have one
simple pole for each pair of positive integers $m$ and $n$ ($n>1$) at
$\hat c=\hat c_{m,n}(\Delta)$, where
\begin{align}
&\hat c_{m,n}=5+2(T_{m,n}+T_{m,n}^{-1}),
\label{Cmn}
\\
&T_{m,n}=\frac{1-4\Delta-mn+\sqrt{[(mn-1)+4\Delta]^2-(m^2-1)(n^2-1)}}{n^2-1}.
\label{Tmn}
\end{align}
Hence, the residues at the poles of the conformal blocks as functions of the
central charge $\hat c$ are also completely determined:
\begin{equation}
\begin{aligned}
\operatorname*{res}_{\hat c=\hat c_{m,n}}
\mathcal{F}_{\text{e}}(\hat c,\Delta,x)&=
\widehat{B}_{m,n}^{\text{(e)}}(\Delta)\begin{cases}
\mathcal{F}_{\text{e}}(\hat c,\Delta+mn/2,x)&\text{for }m,n\text{ even},\\
\mathcal{F}_{\text{o}}(\hat c,\Delta+mn/2,x)&\text{for }m,n\text{ odd},
\end{cases}
\\
\operatorname*{res}_{\hat c=\hat c_{m,n}}
\mathcal{F}_{\text{o}}(\hat c,\Delta,x)&=
\widehat{B}_{m,n}^{\text{(o)}}(\Delta)\begin{cases}
\mathcal{F}_{\text{o}}(\hat c,\Delta+mn/2,x)&\text{for }m,n\text{ even},\\
\mathcal{F}_{\text{e}}(\hat c,\Delta+mn/2,x)&\text{for }m,n\text{ odd},
\end{cases}
\end{aligned}
\label{rescF}
\end{equation}
where
\begin{equation}
\widehat{B}_{m,n}^{\text{(e,o)}}(\Delta)=
B_{m,n}^{\text{(e.o)}}(\Delta)\frac{16(T_{m,n}(\Delta)-T_{m,n}^{-1}(\Delta))}
{(n^2-1)T_{m,n}(\Delta)-(m^2-1)T_{m,n}^{-1}(\Delta)}.
\label{Bmn}
\end{equation}
Relations~\eqref{chain} are simplified for $c=\infty$ and can be solved
explicitly:
\begin{align}
&\mathcal{C}(N,\Delta)=\begin{cases}
\dfrac{(\Delta+\Delta_1-\Delta_2)_N}{N!(2\Delta)_N}G_{-1/2}^{2N},&
N\text{ integer},\\
\dfrac{(\Delta+1/2+\Delta_1-\Delta_2)_{N-1/2}}
{(N-1/2)!(2\Delta+1)_{N-1/2}}G_{-1/2}^{2N},&
N\text{ half-integer},\end{cases}
\\
&\widetilde{\mathcal{C}}(N,\Delta)=\begin{cases}
\dfrac{(\Delta+1/2+\Delta_1-\Delta_2)_N}{N!(2\Delta)_N}G_{-1/2}^{2N},&
N\text{ integer},\\
\dfrac{(\Delta+\Delta_1-\Delta_2)_{N+1/2}}
{(N-1/2)!(2\Delta)_{N+1/2}}G_{-1/2}^{2N},&N\text{ half-integer},\end{cases}
\end{align}
where $(x)_k=x(x+1)\cdots(x+k-1)$. With
\begin{equation}
G_{1/2}^{2N}G_{-1/2}^{2N}\,V_{\Delta}=\begin{cases}
(2 \Delta)_NN!\,V_{\Delta},&N\text{ integer},\\
(2 \Delta)_{N+1/2}(N-1/2)!\,V_{\Delta},&N\text{ half-integer},\end{cases}
\end{equation}
taken into account, this leads to expressions for the asymptotic values of
$\mathcal{F}_{\text{e}}$ and $\mathcal{F}_{\text{o}}$ in terms of
hypergeometric functions:
\begin{equation}
\begin{aligned}
\mathcal{F}_{\text{e}}(\hat c=\infty,\Delta,x)={}&f_{\text{e}}(\Delta,x)
\\
={}&x^{\Delta-\Delta_1-\Delta_2-1/2}
{}_2F_1(\Delta+1/2+\Delta_1-\Delta_2,\Delta+\Delta_3-\Delta_4,2\Delta,x),
\\
\mathcal{F}_{\text{o}}(\hat c=\infty,\Delta,x)={}&f_{\text{o}}(\Delta,x)
\\
={}&\frac{\Delta+\Delta_1-\Delta_2}{2\Delta}x^{\Delta-\Delta_1-\Delta_2}
\\
&{}\times{}_2F_1(\Delta+\Delta_1-\Delta_2+1,\Delta+
\Delta_3-\Delta_4+1/2,2\Delta+1,x).
\end{aligned}
\end{equation}
It is hence clear that we can write the following relations for the conformal
blocks $\mathcal{F}_{\text{e}}$ and $\mathcal{F}_{\text{o}}$:
\begin{align}
\mathcal{F}_{\text{e,o}}(\hat c,\Delta,x)=
f_{\text{e,o}}(\Delta,x)&{}+\sum_{m,n\text{ even}}
\frac{\widehat{B}_{m,n}^{\text{(e,o)}}(\Delta)}{\hat c-\hat c_{m,n}(\Delta)}
\mathcal{F}_{\text{e,o}}(\hat c_{m,n},\Delta+mn/2,x)
\nonumber
\\
&{}+\sum_{\substack{m,n\text{ odd}\\m>1}}
\frac{\widehat{B}_{m,n}^{\text{(o,e)}}(\Delta)}{\hat c-\hat c_{m,n}(\Delta)}
\mathcal{F}_{\text{o,e}}(\hat c_{m,n},\Delta+mn/2,x).
\label{crec}
\end{align}
We can expand $\mathcal{F}_{\text{e}}$ and $\mathcal{F}_{\text{o}}$ in $x$ by
iterating Eqs.~\eqref{crec}:
\begin{equation}
\begin{aligned}
&\mathcal{F}_{\text{e}}(\hat c,\Delta,x)=x^{\Delta-\Delta_1-\Delta_2-1/2}
\sum_{k=0}^{\infty}\mathcal{F}_{\text{e}}^{(k)}x^k,
\\
&\mathcal{F}_{\text{o}}(\hat c,\Delta,x)=x^{\Delta-\Delta_1-\Delta_2}
\sum_{k=0}^{\infty}\mathcal{F}_{\text{o}}^{(k)}x^k.
\end{aligned}
\end{equation}
Using recursive relations~\eqref{crec}, we easily find the first few terms of
the series expansions:
{\allowdisplaybreaks
\begin{align}
\mathcal{F}_{\text{e}}^{(0)}={}&1,
\label{F0}
\\
\mathcal{F}_{\text{e}}^{(1)}={}&
(1/2+\Delta+\Delta_1-\Delta_2)(\Delta+\Delta_3-\Delta_4)(2\Delta)^{-1},
\\
\mathcal{F}_{\text{e}}^{(2)}={}&(1/2+\Delta+\Delta_1-\Delta_2)
(3/2+\Delta+\Delta_1-\Delta_2)(\Delta+\Delta_3-\Delta_4)
\nonumber
\\*
&{}\times(1+\Delta+\Delta_3-\Delta_4)\bigl(4\Delta(1+2\Delta)\bigr)^{-1}
\nonumber
\\*
&{}+\bigl(\bigl(4\Delta^2+\Delta(4+8\Delta_1+8\Delta_2)-
3(4\Delta_1^2+(1-2\Delta_2)^2-4\Delta_1(1+2\Delta_2))\bigr)
\nonumber
\\*
&{}\times\bigl(\Delta^2-3(\Delta_3-\Delta_4)^2+
2\Delta(\Delta_3+\Delta_4)\bigr)\bigr)
\bigl(24\Delta(3+2\Delta)(-1+c+16\Delta/3)\bigr)^{-1}
\nonumber
\\*
&{}-(\Delta_1-\Delta_2)(3/2+\Delta+\Delta_1-\Delta_2)
\nonumber
\\*
&{}\times\bigl(\Delta_3-2\Delta_3^2+\Delta_4+
4\Delta_3\Delta_4-2\Delta_4^2+\Delta(-1+2\Delta_3+2\Delta_4)\bigr)
\nonumber
\\*
&{}\times\bigl((3/2+\Delta)(1+2\Delta)^2(c-2\Delta(3-2\Delta)
(1+2\Delta)^{-1})\bigr)^{-1},
\\*
\mathcal{F}_{\text{o}}^{(0)}={}&(\Delta+\Delta_1-\Delta_2)(2\Delta)^{-1},
\\*
\mathcal{F}_{\text{o}}^{(1)}={}&
2(\Delta_1-2\Delta_1^2+\Delta_2+4\Delta_1\Delta_2-2\Delta_2^2+
\Delta(2\Delta_1+2\Delta_2-1))(\Delta_4-\Delta_3)
\nonumber
\\*
&{}\times\bigl((1+2\Delta)^2(c-2\Delta(3-2\Delta)(1+2\Delta)^{-1})\bigr)^{-1}
\nonumber
\\*
&{}+(1+\Delta+\Delta_1-\Delta_2)(\Delta+\Delta_1-\Delta_2)
(1/2+\Delta+\Delta_3-\Delta_4)
\nonumber
\\*
&{}\times\bigl(2\Delta(1+2\Delta)\bigr)^{-1}.
\label{F1}
\end{align}}%
Further coefficients are also accessible as a solution of~\eqref{crec}. But
it turns out that another recursive representation for the conformal blocks
is more convenient from the practical standpoint. We consider it in the next
section.

\section{q-Recursion construction for the four-point correlation function}

We can use the analytic properties of the superconformal blocks in $\Delta$,
described in Sec.~4, to write a more convenient ``elliptic" recursive
representation. For this, we must know the asymptotic behavior of the
superconformal blocks in the limit $\Delta\to\infty$. At the moment, we are
not able to derive this behavior from general principles, but with
differential equation~\eqref{g2} in hand, we can calculate the series
expansions of the degenerate superconformal blocks up to a high order in $x$.
Based on this calculation, we conclude that the asymptotic behavior is
similar to the asymptotic behavior in the previous case~\cite{SUSYboots}:
\begin{equation}
\begin{aligned}
\mathcal{F}_{\text{e}}\left(\begin{array}[c]{cc}%
\hat\Delta_1&\Delta_3\\ \Delta_2&\Delta_4\end{array}
\biggl|\Delta\biggr|x\right)={}&(16q)^{\Delta-Q^2/8}\frac
{x^{Q^2/8-(\Delta_1+1/2)-\Delta_2}(1-x)^{Q^2/8-(\Delta_1+1/2)-\Delta_3}}
{\theta_3^{2+4\sum_{i=1}^4\Delta_i-3Q^2/2}(q)}
\\
&{}\times H_{\text{e}}\left(\begin{array}[c]{cc}%
\hat\lambda_1&\lambda_3\\ \lambda_2&\lambda_4\end{array}
\biggl|\Delta\biggr|q\right),
\\
\mathcal{F}_{\text{o}}\left(\begin{array}[c]{cc}%
\hat\Delta_1&\Delta_3\\ \Delta_2&\Delta_4\end{array}
\biggl|\Delta\biggr|x\right)={}&(16q)^{\Delta-Q^2/8}\frac
{x^{Q^2/8-(\Delta_1+1/2)-\Delta_2}(1-x)^{Q^2/8-(\Delta_1+1/2)-\Delta_3}}
{\theta_3^{2+4\sum_{i=1}^4\Delta_i-3Q^2/2}(q)}
\\
&{}\times H_{\text{o}}\left(\begin{array}[c]{cc}%
\hat\lambda_1&\lambda_3\\ \lambda_2&\lambda_4\end{array}
\biggl|\Delta\biggr|q\right),
\end{aligned}
\label{FH}
\end{equation}
where we introduce the ``elliptic" blocks $H_{\text{e,o}}(\Delta,q)$. The
elliptic parameter is $q=\exp(i\pi\tau)$, where
\begin{equation}
\tau=i\frac{K(1-x)}{K(x)}.
\label{tau}
\end{equation}
The asymptotic behavior of $H_{\text{e,o}}$ is independent of the internal
conformal dimension $\Delta$:
\begin{equation}
\begin{aligned}
H_{\text{e}}(\Delta,q)&=\theta_3(q^2)+O(\Delta^{-1}),
\\
H_{\text{o}}(\Delta,q)&=\theta_2(q^2)+O(\Delta^{-1}),
\end{aligned}
\label{Hass}
\end{equation}
and
\begin{equation}
\theta_3(q)=\sum_{n=-\infty}^{\infty}q^{n^2},\qquad
\theta_2(q)=\sum_{n=-\infty}^{\infty}q^{(n+1/2)^2}.
\label{theta}
\end{equation}
The elliptic blocks satisfy the relations
\begin{equation}
\begin{aligned}
H_{\text{e}}(\Delta,q)=\theta_3(q^2)&{}+\sum_{m,n\text{ even}}
\frac{q^{mn/2}R_{m,n}^{\text{(e)}}}{\Delta-\Delta_{m,n}}
H_{\text{e}}(\Delta_{m,-n},q)
\\
&{}+\sum_{mn\text{ odd}}\frac{q^{mn/2}R_{m,n}^{\text{(o)}}}
{\Delta-\Delta_{m,n}}H_{\text{o}}(\Delta_{m,-n},q),
\\
H_{\text{o}}(\Delta,q)=\theta_2(q^2)&{}+\sum_{m,n\text{ even}}
\frac{q^{mn/2}R_{m,n}^{\text{(o)}}}{\Delta-\Delta_{m,n}}
H_{\text{o}}(\Delta_{m,-n},q)
\\
&{}+\sum_{m,n\text{ odd}}\frac{q^{mn/2}R_{m,n}^{\text{(e)}}}
{\Delta-\Delta_{m,n}}H_{\text{e}}(\Delta_{m,-n},q),
\end{aligned}
\label{Hrec}
\end{equation}
where the residues are simply
\begin{equation}
R_{m,n}^{\text{(e,o)}}=
\frac{P_{m,n}^{\text{(o,e)}}(\lambda_1+\lambda_2)
P_{m,n}^{\text{(o,e)}}(\lambda_1-\lambda_2)
P_{m,n}^{\text{(e,o)}}(\lambda_3+\lambda_4)
P_{m,n}^{\text{(e,o)}}(\lambda_3-\lambda_4)}{r_{m,n}'}.
\label{Reo}
\end{equation}
Relations~\eqref{Hrec} allow recursively evaluating the series expansions of
the elliptic blocks in $q$,
\begin{equation}
H_{\text{e,o}}(\Delta,q)=\sum_{N\in\mathcal{Z},\mathcal{Z}/2}q^Nh_{\text{e,o}}^{(N)}(\Delta),
\label{hN}
\end{equation}
where again $H_{\text{e}}(\Delta,q)$ expands in nonnegative integer powers of
$q$ and $H_{\text{o}}(\Delta,q)$ expands in positive half-integer ones.
Relations~\eqref{Hrec} give
\begin{equation}
h_{\text{e,o}}^{(N)}(\Delta)=
\eta_{\text{e,o}}^{(N)}+\sum_{m,n\text{ even}}^{mn/2\le N}
\frac{R_{m,n}^{\text{(e,o)}}h_{\text{e,o}}^{(N-mn/2)}(\Delta_{m,-n})}
{\Delta-\Delta_{m,n}}+\sum_{m,n\text{ odd}}^{mn/2\le N}
\frac{R_{m,n}^{\text{(o,e)}}h_{\text{o,e}}^{(N-mn/2)}(\Delta_{m,-n})}
{\Delta-\Delta_{m,n}},
\label{HNrc}
\end{equation}
where $\eta_{\text{e}}^{(N)}$ and $\eta_{\text{o}}^{(N)}$ are coefficients
in the $q$-expansion of the corresponding asymptotic forms~\eqref{Hass}. The
power series in $q$ for the ``elliptic" blocks converge for $|q|<1$, i.e.,
on the whole covering of the $x$ plane with three punctures. With structure
constants~\eqref{C3} and the superconformal blocks known, we can now write
the universal construction for four-point function~\eqref{WVVV}:
\begin{align}
&\langle W_{a_1}(x)V_{a_2}(0)V_{a_3}(1)V_{a_4}(\infty)\rangle
\nonumber
\\
&\qquad=\frac{(x\bar x)^{Q^2/8-(\Delta_1+1/2)-\Delta_2}
[(1-x)(1-\bar x)]^{Q^2/8-(\Delta_1+1/2)-\Delta_3}}
{[\theta_3(q)\theta_3(\bar q)]^{2+4\sum_i\Delta_i-3Q^2/2}}
\mathcal{H}\left(\left.\begin{array}[c]{cc}%
\hat a_1&a_3\\a_2&a_4\end{array}\right|\tau,\bar\tau\right),
\end{align}
where
\begin{align}
\mathcal{H}\left(\left.\begin{array}[c]{cc}%
\hat a_1&a_3\\a_2&a_4\end{array}\right|\tau,\bar\tau\right)=
\int\frac{dP}{4\pi}|16q|^{P^2}\Bigl[&\mathbb{\widetilde C}_{a_1,a_2}^{Q/2+iP}
\mathbb{C}_{a_3,a_4}^{Q/2-iP}H_{\text{e}}(\Delta_P,q)
H_{\text{e}}(\Delta_P,\bar q)
\nonumber
\\
&{}+\mathbb{C}_{a_1,a_2}^{Q/2+iP}\mathbb{\widetilde C}_{a_3,a_4}^{Q/2-iP}
H_{\text{o}}(\Delta_P,q)H_{\text{o}}(\Delta_P,\bar q)\Bigr]
\label{gtau}
\end{align}
and $\Delta_P=Q^2/8+P^2/2$. The crossing symmetry leads to the nontrivial
relation in terms of the auxiliary function $\mathcal{H}$
\begin{equation}
\mathcal{H}\left(\left.\begin{array}[c]{cc}%
\hat a_1&a_3\\a_2&a_4\end{array}\right|\tau,\bar\tau\right)=
(\tau\bar\tau)^{3Q^2/4-2\sum_i\Delta_i-2}
\mathcal{H}\left(\left.\begin{array}[c]{cc}%
\hat a_1&a_2\\a_3&a_4\end{array}\right|-\frac1\tau,-\frac1{\bar\tau}\right).
\label{c2}
\end{equation}
Using the numerical algorithm based on the elliptic representation of the
blocks, we have verified this relation for the external parameters
$a_1=a_2=a_3=a_4=Q/2$ and for different values of the parameters $b$ and
$\tau$. Taking terms up to the order $q^6$ in the power series for
$H_{\text{e,o}}(q)$ into account, we obtain a wide range of distances where
the results differ in the fifth digit, and the convergence improves as the
expansion order increases.

\section{Discussion}

We have analyzed the NS sector of the $N{=}1$ supersymmetric Liouville field
theory. We obtained  the four-point correlation function $\langle WVVV\rangle$, involving one ``top" component $W$ of a primary supermultiplet, while the others are ``bottom" components. This is one of the four basic spinless four-point correlation
functions in the theory. Another correlation function $\langle VVVV\rangle$, involving only the ``bottom'' components $V$, was recently considered in~\cite{SUSYboots}.
The other two basic correlation functions $\langle VVWV\rangle$ and $\langle WVWV\rangle$, are constructed in term of the corresponding
superconformal blocks~\eqref{ConfBlockDef1a},\eqref{ConfBlockDef3} and~\eqref{ConfBlockDef1b},\eqref{ConfBlockDef3a} in the same way as it was discussed in this paper
for the correlation function $\langle WVVV\rangle$.
All other spinless four-point functions involving top and bottom
supermultiplet components are expressed in term of these four functions using
superprojective Ward identities.

Our analysis generalizes the approach developed quite long ago
in~\cite{cblock,dblock} for the ordinary CFT. The construction involves the
corresponding four-point superconformal blocks, for which an efficient
recursive calculation procedure was offered. This approach is based on
knowing the asymptotic properties of the conformal blocks in $\Delta$. A
straightforward derivation requires a separate investigation and is the
subject of a future report. But with the additional assumption about the
general structure of this asymptotic behavior, we could recover its explicit
form by analyzing the degenerate case and the corresponding differential
equation. We verified our assumption up to the order $q^{20}$. Remarkably,
the asymptotic behavior of the superconformal blocks is the same as that
considered in~\cite{SUSYboots} for another correlation function. Apparently,
all four basic correlation functions have the same asymptotic properties.

Of course, to complete the analysis of the four-point sector in the super
Liouville field theory, the analogous research should be done for the Ramond
sector of the theory. We hope that the methods discussed here will help in
considering minimal supergravity, where, in contrast to the bosonic analogue
for which the matrix model approach is known, it will be the only way to
calculate numerically.

\medskip

\textbf{Acknowledgments.}
The author sincerely thanks G.~Delfino and G.~Mussardo for their hospitality
at SISSA and for encouraging the interest in this work. He is grateful to
A.~Belavin and Al.~Zamolodchikov for the valuable discussions. This work was
supported by the MIUR program ``Quantum Field Theory and Statistical
Mechanics in Low Dimensions" and by the Russian Foundation for Basic Research
(Grant No.~05-01-01007).

\appendix

\section{Derivation of the differential equations for the correlation
functions $\langle\widehat V_{-b}\widehat V_1\widehat V_2\widehat V_3\rangle$}

First, we want to derive the supersymmetric form of the differential
equations that follow from decoupling condition~\eqref{singvec}. Temporarily
ignoring the existence of the ``left" subalgebra $\overline{SVir}$ and
completely omitting the dependence of the fields on $\bar z$ and
$\bar\theta$, we write
\begin{equation}
\widehat V_{-b}(z,\theta)=(1+\theta G_{-1/2})V_{-b}(z).
\end{equation}
It is straightforward to verify the identity
\begin{equation}
\begin{aligned}
(1- \theta G_{-1/2})(G_{-1/2}^3+b^2 G_{-3/2})V_{-b}=
\biggl(\frac d{dz}D+b^2(G_{-3/2}+2\theta L_{-2})\biggr)\widehat V_{-b}=0,
\label{nulvec}
\end{aligned}
\end{equation}
where the supercovariant derivative $D$ is defined as
\begin{equation}
D=\frac d{d\theta}-\theta\frac d{dz}.
\end{equation}
Inside the correlation function, we can rewrite~\eqref{nulvec} more
universally. In particular, for the four-point function, we have
\begin{equation}
\left\langle\biggl(\frac d{dz}D+b^2\int_{C_z}\frac{du}{u-z}
\widehat S(u,\theta)\biggr)
\widehat V_{-b}\widehat V_1\widehat V_2\widehat V_3\right\rangle=0,
\end{equation}
where $\widehat S(u,\theta)=S(u)-2\theta T(u)$ is ``super stress--energy
tensor" generating superalgebra~\eqref{The algebra} and the contour $C_z$
emcompasses the point $z$. We note that the ``odd" argument of $S$ is the
same as that of the field $V_{-b}$.

The next standard step is to use supercurrent Ward identities. We deform the
integration contour $C_z=-\sum_kC_k$ (permutations of the superfields do not
affect the sign of the whole expression), and taking the OPE written in the
supersymmetric form
\begin{equation}
\widehat S(u,\theta')\widehat V_{\Delta}(z,\theta)=
\frac{2\Delta(\theta-\theta'){\widehat V_{\Delta}(z,\theta)}}{(u-z)^2}+
\frac{2(\theta-\theta')\partial\widehat V_{-b}(z,\theta)}
{u-z+\theta'\theta}+\frac{D\widehat V_{-b}(z,\theta)}{u-z+\theta'\theta}+
\text{reg}
\end{equation}
into account, we immediately derive the supersymmetric partial differential
equation for $\langle\widehat V_{-b}\widehat V_1\widehat V_2
\widehat V_3\rangle$:
\begin{equation}
\left\{-b^{-2}\partial_4D_4+\sum_{i=1}^3\left\{\frac{2\Delta_i}{z_{4i}^2}
\theta_{4i}+\frac1{z_{4i}}\biggl[2\theta_{4i}\partial_i-D_i\biggr]\right\}
\right\}\langle\widehat V_{-b}(z_4)\widehat V_1(z_1)\widehat V_2(z_2)
\widehat V_3(z_3)\rangle=0,
\label{difur0}
\end{equation}
where the arguments are indexed for future purposes. In accordance
with~\eqref{anzac}, we then introduce the function
\begin{equation}
g(z,\tau_1,\tau_2)=g_0(z)+g_1(z)\tau_1+g_2(z)\tau_2+g_3(z)\tau_1\tau_2
\label{g0123new}
\end{equation}
related to the four-point correlation function by
\begin{equation}
\langle\widehat V_{-b}(z_4)\widehat V_1(z_1)\widehat V_2(z_2)
\widehat V_3(z_3)\rangle=z_{34}^{\gamma_{34}}z_{13}^{\gamma_{13}}
z_{23}^{\gamma_{23}}z_{12}^{\gamma_{12}}g(z,\tau_1,\tau_2).
\label{anzac2}
\end{equation}
Here, we again neglect the antiholomorphic dependence of the correlation
function. After the field arguments are renumbered, the cross ratio is
$z=z_{41}z_{23}/z_{43}z_{21}$, and the exponents are given by
\begin{equation}
\begin{aligned}
&\gamma_{34}=-2\Delta_4,
\\
&\gamma_{13}=-\Delta_1-\Delta_3+\Delta_4+\Delta_2,
\\
&\gamma_{23}=-\Delta_2-\Delta_3+\Delta_4+\Delta_1,
\\
&\gamma_{12}=-\Delta_4-\Delta_1-\Delta_2+\Delta_3.
\end{aligned}
\end{equation}
We derive the equation for the function $g$ from~\eqref{difur0}:
\begin{align}
\biggl[-b^{-2}\partial_4D_4-b^{-2}\frac{\gamma_{34}}{z_{34}}
\biggl(\theta_{43}\partial_4-D_4\biggr)+
\sum_{i=1}^3\biggl[\frac{2\Delta_i}{z_{4i}^2}\theta_{4i}+
\frac1{z_{4i}}\biggl(2\theta_{4i}\partial_i-D_i\biggr)\biggr]&
\nonumber
\\
{}+\frac{\gamma_{12}(\theta_{41}+\theta_{42})}{z_{41}z_{42}}+
\frac{\gamma_{13}(\theta_{41}+\theta_{43})}{z_{41}z_{43}}+
\frac{\gamma_{23}(\theta_{42}+\theta_{43})}{z_{42}z_{43}}+
\frac{\gamma_{34}\theta_{43}}{z_{43}^2}&\biggr]g=0,
\label{difur}
\end{align}
For the functions $g_i$, PDE~\eqref{difur} reduces to an ordinary
differential equation in $z$ (of the second order) of the form
\begin{equation}
A_ig_i''+B_ig_i'+C_ig_i=0,
\label{difur1}
\end{equation}
where the explicit expressions for $A_i$, $B_i$, and $C_i$ are
\begin{equation}
\begin{aligned}
&A_0=-b^{-2}D_4z\partial_4z,
\\
&A_1=-b^{-2}D_4z\partial_4z\tau_1,
\\
&A_2=-b^{-2}D_4z\partial_4z\tau_2,
\\
&A_3=-b^{-2}D_4z\partial_4z\tau_1\tau_2,
\end{aligned}
\label{A}
\end{equation}
\begin{equation}
\begin{aligned}
&B_0=-b^{-2}\partial_4D_4z-b^{-2}
\frac{\gamma_{34}}{z_{34}}(\theta_{43}\partial_4z-D_4z)+
\sum_{i=1}^3\frac1{z_{4i}}(2\theta_{4i}\partial_iz-D_iz),
\\
&B_1=B_0\tau_1,
\\
&B_2=B_0\tau_2-b^{-2}D_4z\partial_4\tau_2-b^{-2}\partial_4zD_4\tau_2,
\\
&B_3=B_0\tau_1\tau_2+(b^{-2}D_4z\partial_4\tau_2+
b^{-2}\partial_4zD_4\tau_2)\tau_1,
\end{aligned}
\label{B}
\end{equation}
\begin{equation}
\begin{aligned}
C_0={}&\sum_{i=1}^3\frac{2\Delta_i}{z_{4i}^2}\theta_{4i}+
\frac{\gamma_{12}(\theta_{41}+\theta_{42})}{z_{41}z_{42}}+
\frac{\gamma_{13}(\theta_{41}+\theta_{43})}{z_{41}z_{43}}+
\frac{\gamma_{23}(\theta_{42}+\theta_{43})}{z_{42}z_{43}}+
\frac{\gamma_{34}\theta_{43}}{z_{43}^2},
\\
C_1={}&\sum_{i=1}^3\frac1{z_{4i}}(2\theta_{4i}\partial_i\tau_1-D_i\tau_1)+
C_0 \tau_1,
\\
C_2={}&-b^{-2}\partial_4D_4\tau_2-b^{-2}\frac{\gamma_{34}}{z_{34}}
(\theta_{43}\partial_4\tau_2-D_4\tau_2)+
\sum_{i=1}^3\frac1{z_{4i}}(2\theta_{4i}\partial_i\tau_2-D_i\tau_2)+C_0\tau_2,
\\
C_3={}&b^{-2}\tau_1\partial_4D_4\tau_2+b^{-2}\frac{\gamma_{34}}{z_{34}}
(\theta_{43}\partial_4\tau_2-D_4\tau_2)\tau_1
\\
&{}+\sum_{i=1}^3\frac1{z_{4i}}[(2\theta_{4i}\partial_i\tau_1-D_i\tau_1)\tau_2-
(2\theta_{4i}\partial_i\tau_2-D_i\tau_2)\tau_1]+C_0\tau_1\tau_2.
\end{aligned}
\label{C}
\end{equation}
To simplify the calculations, we choose the ``odd" superprojective invariants
slightly differently here than in the main text in~\eqref{params1}:
$\tau_1=\theta_{213}$ and $\tau_2=\theta_{214}$.

Our next task is to find explicit expressions for each term
in~\eqref{A},~\eqref{B}, and~\eqref{C}. We again use superprojective
invariance, fixing the particular choice
\begin{equation}
\begin{alignedat}{3}
&\theta_1=0,&\qquad&\theta_2=0,&\qquad&\theta_3=R\eta,
\\
&x_1=0,&\qquad&x_2=1,&\qquad&x_3=R,\qquad x_4=x,
\end{alignedat}
\end{equation}
where $R\to\infty$. We calculate the coefficients $A_i$, $B_i$, and $C_i$ in
terms of $z$, $\tau_1$, and $\tau_2$. In particular,
\begin{equation}
\begin{aligned}
&\eta=\tau_1,
\\
&\theta_4=z^{1/2}(z-1)^{1/2}\tau_2,
\\
&x=z[1- z^{1/2}(z-1)^{1/2}\tau_1\tau_2].
\end{aligned}
\label{z-x}
\end{equation}
In the same manner, we write all the derivatives explicitly in terms of $z$,
$\tau_1$, and $\tau_2$, which results in the expressions for the coefficients
$A_i$, $B_i$, and $C_i$:
\begin{equation}
\begin{aligned}
&A_0=-b^{-2}z\tau_1+b^{-2}z^{1/2}(z-1)^{1/2}\tau_2,
\\
&A_1=-b^{-2}z^{1/2}(z-1)^{1/2}\tau_1\tau_2,
\\
&A_2=-b^{-2}z\tau_1\tau_2,
\\
&A_3=0,
\end{aligned}
\end{equation}
\begin{equation}
\begin{aligned}
&B_0=\biggl[\frac z{z-1}-b^{-2}(1-\gamma_{34})\biggr]\tau_1+
\frac{1-3z}{z(z-1)}z^{1/2}(z-1)^{1/2}\tau_2,
\\
&B_1=-\frac{1-3z}{z(z-1)}z^{1/2}(z-1)^{1/2}\tau_1\tau_2,
\\
&B_2=-\frac{b^{-2}}{z^{1/2}(z-1)^{1/2}}+\biggl[\frac z{z-1}+
b^{-2}\biggl(\frac{3z-2}{z-1}+\gamma_{34}-1\biggr)\biggr]\tau_1\tau_2,
\\
&B_3=\frac{b^{-2}}{z^{1/2}(z-1)^{1/2}}\tau_1,
\end{aligned}
\end{equation}
\begin{equation}
\begin{aligned}
&C_0=\biggl(\frac{\gamma_{13}}{z}+\frac{\gamma_{23}}{z-1}\biggr)\tau_1+
\biggl(\frac{2\Delta_1}{z^2}+\frac{2\Delta_2}{(z-1)^2}+
\frac{2\gamma_{12}}{z(z-1)}\biggr)z^{1/2}(z-1)^{1/2}\tau_2,
\\
&C_1=\frac1{z(z-1)}-\bigg(\frac{2\Delta_1}{z^2}+\frac{2\Delta_2}{(z-1)^2}+
\frac{2\gamma_{12}(z-1)+3z-2}{z(z-1)^2}\biggr)z^{1/2}(z-1)^{1/2}\tau_1\tau_2,
\\
&C_2=\biggl(1+\frac{b^{-2}}2\biggr)\frac{2z-1}{z^{3/2}(z-1)^{3/2}}+
\biggl[\frac{\gamma_{13}}z+\frac{\gamma_{23}}{z-1}-
\frac{(2+b^{-2})(12z^2-14z+5)}{4z(z-1)^2}\biggr]\tau_1\tau_2,
\\
&C_3=[2+b^{-2}]\frac{1-2z}{2z^{3/2}(z-1)^{3/2}}\tau_1+\frac1{z(z-1)}\tau_2.
\end{aligned}
\end{equation}
Substituting in differential equation~\eqref{difur1} and separating linearly
independent contributions, we finally obtain two independent sets of ordinary
linear differential equations. The first set, involving $g_0$ and $g_2$, is
obtained by collecting the terms with $\tau_1$ and $\tau_2$:
\begin{equation}
\begin{aligned}
-b^{-2}zg_0''+\frac{3z-2}{z-1}g_0'+\frac{b^{-2}}{z^{1/2}(z-1)^{1/2}}g_3'+
\biggl[\frac{\gamma_{13}}z+\frac{\gamma_{23}}{z-1}\biggr]g_0&
\\
{}+\frac{(2+b^{-2})(1-2z)}{2z^{3/2}(z-1)^{3/2}}g_3&=0 ,
\\
b^{-2}g_0''+\frac{1-3z}{z(z-1)}g_0'+\biggl[\frac{2\Delta_1}{z^2}+
\frac{2\Delta_2}{(z-1)^2}+\frac{2\gamma_{12}}{z(z-1)}\biggr]g_0+
\frac1{z^{3/2}(z-1)^{3/2}}g_3&=0.
\end{aligned}
\label{g0g3}
\end{equation}
The second set of equations, coming from the terms that are even in $\tau_1$
and $\tau_2$, involve $g_1$ and $g_2$:
\begin{equation}
\begin{aligned}
-b^{-2}z^{1/2}(z-1)^{1/2}g_1''-b^{-2}zg_2'+
\frac{3z-1}{z(z-1)}z^{1/2}(z-1)^{1/2}g_1'+(1+b^{-2})\frac{3z-2}{z-1}g_2'&
\\
{}-\biggl[\frac{2\Delta_1}{z^2}+\frac{2\Delta_2}{(z-1)^2}+
\frac{2\gamma_{12}(z-1)+3z-2}{z(z-1)^2}\biggr]z^{1/2}(z-1)^{1/2}g_1&
\\
{}+\biggl[\frac{\gamma_{13}}z+\frac{\gamma_{23}}{z-1}-
(2+b^{-2})\frac{12z^2-14z+5}{4z(z-1)^2}\biggr]g_2&=0,
\\
-\frac{b^{-2}}{z^{1/2}(z-1)^{1/2}}g_2'+\frac1{z(z-1)}g_1+
\biggl(1+\frac{b^{-2}}2\biggr)\frac{2z-1}{z^{3/2}(z-1)^{3/2}}g_2&=0.
\end{aligned}
\label{g1g2}
\end{equation}
Finally, we replace $\tau_2$ with $z^{1/2}(z-1)^{1/2}\tau_2$.
With~\eqref{g0123new} taken into account, this is equivalent to replacing
$g_2$ with $z^{-1/2}(z-1)^{-1/2}g_2$ and $g_3$ with
$z^{-1/2}(z-1)^{-1/2}g_3$. This substitution converts~\eqref{g0g3}
and~\eqref{g1g2} to the more transparent forms~\eqref{syst1}
and~\eqref{syst2}.

\end{document}